\documentclass[a4paper,fleqn,a4paper]{cas-dc}
\usepackage{cite}
\usepackage{amsmath,amssymb,amsfonts}
\usepackage{algorithmic}
\usepackage{graphicx,color}
\usepackage{textcomp}
\usepackage{xcolor}

\usepackage{algorithm,algorithmic}
\usepackage{booktabs}
\usepackage{multicol}
\usepackage{multirow}
\usepackage{tabularx}
\usepackage{longtable}
\usepackage{threeparttable}
\usepackage{algorithmic}
\usepackage{graphicx}
\usepackage{textcomp}
\usepackage{xcolor}
\usepackage{balance}
\usepackage{array}
\usepackage{colortbl}
\usepackage[most]{tcolorbox}
\usepackage{paralist} 
\usepackage{pifont}

\usepackage{soul}

\usepackage[resetlabels,labeled]{multibib}
\newcites{S}{Primary Studies}

\AtBeginDocument{\definecolor{tmlcncolor}{cmyk}{0.93,0.59,0.15,0.02}\definecolor{NavyBlue}{RGB}{0,86,125}}

\newcommand{\sectopic}[1]{\par\noindent{\textit{\bfseries #1}}}
 \newcommand{\revv}[1]{\textcolor{black}{#1}}
 \newcommand{\revn}[1]{\textcolor{black}{#1}}

\newcommand\rev[1]{\textcolor{black}{#1}}

\usepackage{hyperref}

\begin{document}
\let\WriteBookmarks\relax
\def\floatpagepagefraction{1}
\def\textpagefraction{.001}

\shorttitle{GDPR in Mobile Apps Research: A Systematic Literature Review}   

\shortauthors{Amaral \textit{et al.}}  

\title[mode=title]{GDPR-Relevant Privacy Concerns in Mobile Apps Research: A Systematic Literature Review}

\author[2]{Orlando Amaral 
Cejas}[orcid=0000-0001-5140-6439]
\cormark[1]
\ead{orlando.amaral-cejas@list.lu}
\credit{Methodology, Validation, Investigation, Data Curation, Writing - Original Draft}

\author[1]{Sallam Abualhaija}[orcid=0000-0001-6095-447X]

\cormark[2]
\fnmark[1]

\ead{sallam.abualhaija@uni.lu}


\credit{Conceptualization, Methodology, Validation, Writing - Review \& Editing, Supervision, Funding acquisition}

\author[1]{Nicolas Sannier}[orcid=0000-0002-4449-5792]

\fnmark[1]

\ead{nicolas.sannier@uni.lu}


\credit{Conceptualization, Methodology, Validation, Writing - Review \& Editing}

\author[1]{Marcello Ceci}[orcid=0000-0003-3800-0906]

\fnmark[1]

\ead{marcello.ceci@uni.lu}


\credit{Writing - Review \& Editing}

\author[1]{Domenico Bianculli}[orcid=0000-0002-4854-685X]

\fnmark[1]

\ead{domenico.bianculli@uni.lu}


\credit{Conceptualization, Writing - Review \& Editing, Supervision, Funding acquisition}

\affiliation[1]{organization={University of Luxembourg},
            country={Luxembourg}}
            
            \affiliation[2]{organization={Luxembourg Institute of 
            Science and Technology},
            	country={Luxembourg}}


\cortext[cor1]{Part of this work was done while the author was affiliated with University of Luxembourg, Luxembourg}
\cortext[cor2]{Corresponding author}

\begin{abstract}
\noindent\textit{Context:} The General Data Protection Regulation (GDPR) is considered as the benchmark in the European Union (EU) for privacy and data protection standards.
Since before its entry into force in 2018, substantial research has been conducted in the \rev{software engineering (SE) literature} investigating the elicitation, representation, and verification of GDPR privacy requirements. 
Software systems deployed anywhere in the world must comply with GDPR as long as they handle personal data of EU residents. Mobile applications (apps) are no different in that regard. With the growing pervasiveness of mobile apps and their increasing demand for personal data, privacy concerns have acquired further interest within the SE community.  Despite the extensive literature on GDPR-relevant privacy concerns in mobile apps, there is no secondary study that describes, analyzes, and categorizes the current focus. Research gaps and persistent challenges are thus left unnoticed.

\textit{Objective:} This article \rev{ aims to provide a comprehensive overview of the existing research on GDPR privacy concerns in the context of mobile apps. }

\textit{Methods:} To do so, we conducted a systematic literature review \rev{of 60 primary studies. }

\textit{Results:} Our findings show that 
\rev{existing studies predominantly address three key GDPR-related privacy concerns: (i) the direct collection of personal data from users, (ii) the sharing of personal data with external entities (e.g., third parties) beyond the mobile apps, and (iii) the analysis of  user consent as a legal basis for collecting personal data.} 

\textit{Conclusion:} 
\rev{Our study highlighted research gaps, calling for further research to better understand: (i) the indirect collection of personal data, e.g., data exposed to mobile apps through, e.g., permission requests, (ii) the impact of legal bases beyond consent and how they may affect the development of mobile apps, and (iii) the required implementation details pertinent to data subject rights.}

\end{abstract}

\begin{keywords}
\textbf{Keywords:} Systematic Literature Review (SLR) \sep Software Engineering (SE) \sep Regulatory Compliance \sep The General Data Protection Regulation (GDPR) \sep Privacy Requirements \sep Mobile Apps
\end{keywords}

\maketitle

\section{Introduction}
\label{sec:introduction}

With the widespread adoption of mobile applications (apps) in various domains, e.g., language learning~\cite{godwin2012}, healthcare~\cite{Fernandez-Aleman2013,martinez2013,milne2020}, and finance~\cite{shaikh2022}, regular sharing of personal data has become the norm. 
\rev{Alongside this growing adoption, users increasingly express} concerns and expectations regarding how their personal data is collected, processed, or shared~\cite{KPMG2024}.
\rev{Simply downloading a mobile app may already trigger the automatic
  sharing of data. } For instance, clicking the ``download'' button can contribute not only to the app's download statistics, but also to reveal the user preferences, which may later be exploited  by, e.g., app market recommender systems.
Such implicit data collection practices can pose a threat to the user's privacy.
Several  studies~\cite{Martinez-Perez2014, 
Morales-Trujillo2019, caramujo2019rsl, Semantha2020, 
Shrivastava2020, Andrade2023} have acknowledged the 
prevalence of such  threats, highlighting that mitigating 
privacy risks in mobile apps requires a deeper understanding 
of GDPR.

In response to these growing privacy concerns, regulations---such as the General Data Protection Regulation (GDPR)~\cite{EU2018} issued by the European Union (EU)---have introduced various privacy requirements that should, if implemented appropriately, guarantee the protection of individuals' personal data throughout the data processing chain. 
In the context to mobile apps, GDPR imposes obligations onto 
development companies, regardless of whether EU-based or 
not, whenever they collect or process personal data of EU 
residents. The GDPR further levies hefty fines on companies 
that fail to comply with such obligations. 
\revv{Among existing privacy regulations, we focus on GDPR 
because of
its global reach and its significant impact
on software development practices 
worldwide~\cite{iwaya2023}, making it one
of the most influential privacy regulations governing the
collection and processing of personal data. Restricting the 
scope
of this study to GDPR also enables a more focused 
analysis of privacy concerns and compliance challenges in the
mobile-app domain.}

GDPR has been extensively investigated in both the legal and 
technological domains. 
\rev{In software engineering (SE), the literature covers} a wide variety of GDPR-related challenges, including modeling (or representing) legal requirements~\cite{Amaral:21,Ghanavati2014}, categorizing and analyzing legal documents~\cite{azeem2024,Amaral:22,Torre:20RE,Bhatia19}, investigating the users' awareness of privacy~\cite{li2022,omoronyia2013}, and assessing privacy concerns in software applications~\cite{Fan:20,Kununka:17}. 
Privacy concerns in mobile apps research have also been investigated to a large extent~\cite{Iwaya2020,Alloghani2020}.

\revn{Although several primary studies have investigated 
specific GDPR-related privacy concerns in mobile applications, 
such as consent (e.g.,~\cite{Tsohou2017}), app permissions 
(e.g.,~\cite{Marsch2021}), and 
data 
sharing with third-party (e.g.,~\cite{samhi2023}), }there is still 
a 
lack of 
secondary 
studies that 
systematically gather and consolidate the diverse body of 
research on mobile applications through the lens of 
fine-grained GDPR-relevant privacy concepts.

In this article, we \textit{\rev{present the results of} a 
systematic literature review (SLR),} \rev{conducted according 
to the best practices and guidelines in the 
literature~\cite{KitchenhamBrereton2013,KitchenhamCharters2007},}
 and 
\revv{\textit{provide a comprehensive 
overview of existing research on GDPR in the context of mobile 
applications.} Specifically, we analyze the GDPR-relevant 
privacy concerns investigated in prior studies, identify the 
concerns that have received substantial attention, and 
highlight 
under-explored concerns that require further research to 
reduce potential compliance risks.}
We analyzed a total of 60 primary studies against an established conceptual model and \rev{a list of principles} that collectively represent  \rev{the most comprehensive set of } privacy-relevant concepts derived from the GDPR provisions. 
The conceptual model we adopted,  proposed by Amaral et al.~\cite{Amaral:21}, consists of 56 \rev{concepts} pertinent to GDPR privacy policies. Examples include: \textit{DATA SUBJECT RIGHT}, \rev{referring to  individuals' rights over their personal data }; \textit{LEGAL BASIS}, referring to the legal basis under which personal data is collected (e.g., through explicit consent obtained from individuals); and \textit{TRANSFER OUTSIDE EUROPE}, encapsulating the necessary mechanisms required for transferring personal data outside the EU.
\rev{Complementing this model, we refer to the set of principles defined in Art. 5 in the GDPR, such as data minimization, purpose limitation, or fairness, transparency, and accountability.}

In this SLR, we primarily \rev{map important facets of existing research in the SE literature against our defined set of GDPR-related privacy concerns}.
\rev{These facets include pursued research objectives, the types of personal data investigated, and the nature of contributions.}
The overarching goal of our study is \textit{to identify the main topics highlighted in the literature and uncover the research gaps that should be further investigated by the community. }

Our study focuses exclusively on GDPR. We thus restrict the 
time span of the primary studies to 2016--2023, considering 
that discussions about GDPR commenced in 2016, before it 
came into force in 2018. Prior to 2016, personal data 
protection in the EU was based on national transpositions of 
the 1995 Directive on personal data protection~\cite{EU95}. 
We cover work across  \rev{various areas of computer science and engineering, such as software engineering and security, as well as different venues}, targeting research studies that investigate GDPR-relevant privacy concerns in mobile apps. Our search process resulted in 484 primary studies that were reduced to 60 relevant studies  to our analysis\citeS{S1,S2,S3,S4,S5,S6,S7,S8,S9,S10,S11,S12,S13,S14,S15,S16,S17,S18,S19,S20,S21,S22,S23,S24,S25,S26,S27,S28,S29,S30,S31,S32,S33,S34,S35,S36,S37,S38,S39,S40,S41,S42,S43,S44,S45,S46,S47,S48,S49,S50,S51,S52,S53,S54,S55,S56,S57,S58,S59,S60}.
To conduct our review,  we extracted information from all 
relevant papers, concerning eight categories, namely (1) the 
GDPR concepts and principles covered in the primary studies; 
(2) the traced personal data; (3) the targeted app stores and 
operating systems; (4) the goals and contributions of the 
primary studies; (5) the applied data analysis methods; (6) the 
proposed solutions and employed technologies; (7) the data 
used; and (8) whether any material is made publicly available.

\rev{\sectopic{Contributions. } 
The paper makes the following four contributions.}

\begin{asparaenum}
  \item 
Our results indicate that existing work focuses on a subset of 
the GDPR-relevant privacy concerns in mobile apps.
Specifically, the following GDPR concerns have been prominently investigated: 
\begin{enumerate}
    \item \rev{Direct collection of personal data from users and the respective categories of collected personal data. }
    \item Sharing of personal data  with third-party libraries, mainly to identify \rev{risks of unintended  disclosure to external entities. }
    \item Mechanisms for obtaining user consent, often forming the legal basis on which personal data is collected.
\end{enumerate}
While such GDPR concepts are indeed important for SE, other concepts in GDPR remain under-explored in the literature, e.g., \rev{the data subject rights over their personal data }. 
In this work, we highlight future research directions aiming at advancing the development of GDPR-compliant mobile apps. 

\item 
\rev{We have identified eight research themes that concentrate current research efforts. These include tracing personal data for detecting potential data leakage, analyzing consent implementation and data minimization principles, checking the consistency between granted permissions, data use, and privacy policies, as well as topics addressing the privacy of minors, users' privacy awareness, and the summarization of privacy policies.}

\item 
\rev{We observed that roughly half of the contributions consist of high-level surveys and addressing broad privacy concerns, while the other half propose technical solutions targeting specific issues.}

\item    
We identified key research gaps and outlined future research directions to advance  SE research on GDPR privacy in mobile apps, by mapping existing work against the same conceptual model and set of principles.
\end{asparaenum}

\rev{\sectopic{Data Availability Statement.} The data supporting our findings, including the data extraction materials and a glossary for the model concepts, is made publicly available in an online annex~\cite{annex}.}

\sectopic{Structure.} 
The remainder of this paper is structured as follows: Section~\ref{sec:background} provides background information. Section~\ref{sec:related} surveys the state of the art. Section~\ref{sec:studyDesign} presents the design of our systematic literature review. Section~\ref{sec:findings} reports on our findings. Section~\ref{sec:threats} discusses threats to validity. Section~\ref{sec:conclusion} concludes the paper.

\section{Background}
\label{sec:background}

\sectopic{GDPR compliance in mobile apps. } 
\rev{Mobile phones and apps have now become prevalent in our digital life. 
From 0.7\% of the global internet traffic in 2009, mobile phones now represent around 60\% of the global internet traffic\footnote{\url{https://web.archive.org/web/20250925150757/https://www.statista.com/statistics/277125/share-of-website-traffic-coming-from-mobile-devices/}}, way more than desktop computers.
Users spend most of their times on their apps, preferring them 
over 
web browsers for their personalized features, convenient use, and performance.
On the other hand, mobile devices are more vulnerable 
due to security challenges, increasing the risk to expose personal data of users through, e.g., the interactions with third party services---as part of the monetization system in mobile apps.
As a consequence, personal data protection  and privacy concerns in
mobile apps are increasingly important topics under GDPR 
compliance. 
}

\sectopic{GDPR Normative Requirements.} 
\rev{GDPR has been widely studied in the literature, where various representations were proposed for capturing and analyzing privacy concerns in the GDPR including principles~\cite{Aljeraisy2021,Tamburri2020} and other requirements, such as processing purposes~\cite{VaneziKKPP20}, consent~\cite{Fan2020}, direct data collection~\cite{Fan2020}, data sharing with third parties~\cite{Nejad2020}, some data subject rights~\cite{Aljeraisy2021,Fan2020,Nejad2020}, data retention~\cite{Nejad2020}, security measures~\cite{Fan2020,Nejad2020}, or contact information~\cite{Fan2020}.}
\rev{GDPR principles are found in the preamble of the regulation, such as paragraph 78  on ``data minimization''  and paragraph 156 on ``data protection by design'', or within the articles, e.g., art. 5 on ``data minimization'' and art. 25 on `data protection by design''.
	It is worth noting that the preamble is considered as 
	non-normative part. 
These principles represent high-level objectives, compared to 
the rights and requirements that are more precise privacy 
concerns stipulated in the normative part of the regulation, 
e.g., art. 12 on ``data subject rights''.} In addition to the rights 
and requirements (elaborated below), this study considers the 
GDPR principles mentioned in art. 5, namely lawfulness, 
fairness, transparency, purpose limitation, data minimization, 
accuracy, storage limitation, integrity, confidentiality, and 
accountability.

\sectopic{Modeling GDPR privacy concerns. } 
In this SLR, we utilize the conceptual model proposed by 
Amaral et al.~\cite{Amaral:21} as a basis for analyzing the 
primary studies.  
Although  the model was
	initially developed to support the completeness assessment 
	of 
	privacy policy, it was created in collaboration with legal
	experts and is grounded in GDPR provisions, particularly 
	those
	contained in Chapters III--V.
Consequently, the model captures
a comprehensive set of GDPR-relevant privacy concepts that
extend beyond privacy-policy analysis and reflect the
underlying data processing practices, obligations, and rights
established by the GDPR. 
\revn{We selected this model motivated by three 
considerations: 
1) \textit{Coverage}: 
To the best 
of our knowledge, it provides a
comprehensive set of privacy concepts specifically tailored to 
the GDPR-related concerns, 
whereas other privacy models proposed in the literature are 
generally broader and more generic 
(e.g.,~\cite{sangaroonsilp2023}). 
2) \textit{Authority and validity: } 
The model is grounded in legal interpretations of GDPR 
requirements and was developed with close involvement of 
legal experts, providing a validated conceptualization of 
privacy concerns under investigation. 
3) \textit{Methodological suitability: } as our goal is to 
systematically synthesize evidence rather than define a new 
taxonomy of privacy concerns, the model offers a stable and 
well-defined analytical framework that supports consistent, 
reproducible, and comparable mapping of the primary studies. 
Therefore, in this
study, we map the primary studies to the model concepts as 
defined in the original model. }

\begin{figure*} 
  \includegraphics[width=\textwidth]{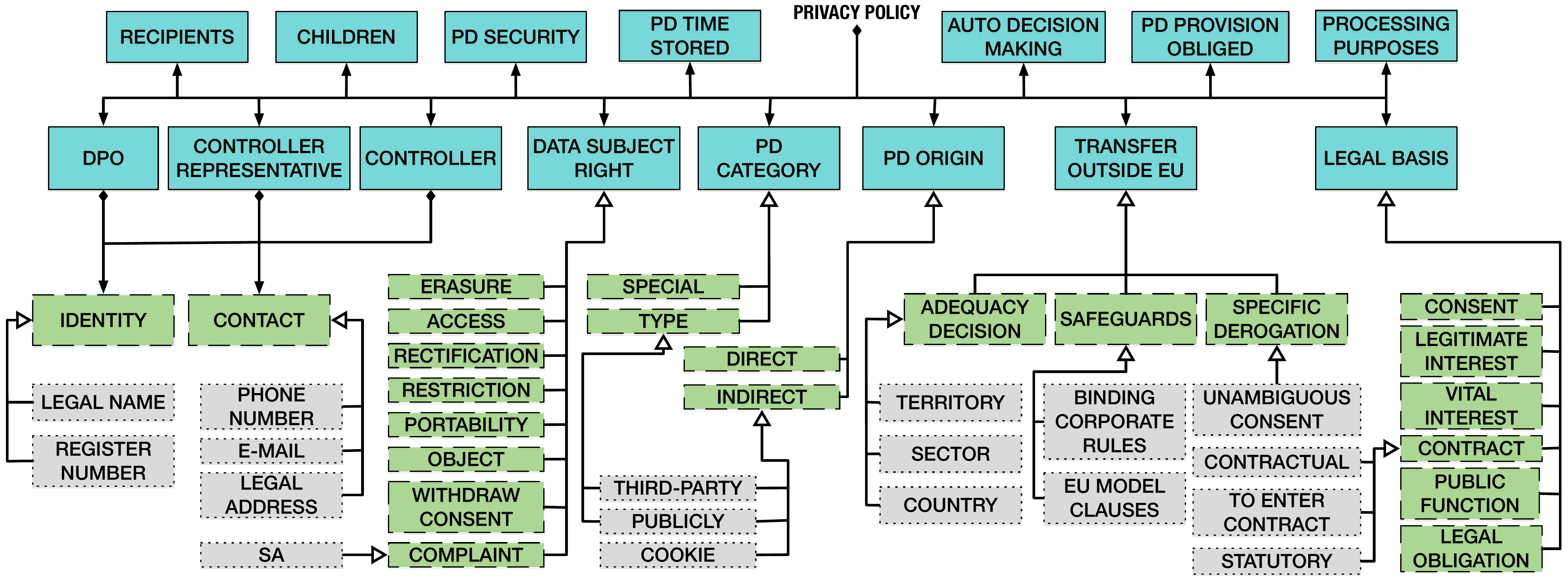}
  \caption{Overview of the \textit{Reference model}~\cite{Amaral:21} Used in this SLR for Identifying GDPR-related Privacy Concepts}
  \label{fig:model}
\end{figure*}

The conceptual model, hereafter referred to as \textit{reference model}, is depicted in Fig.~\ref{fig:model}.
The model is composed of 56 
\rev{concepts} organized in three hierarchical levels.
Level~1, shaded teal-blue (solid-edge boxes), captures high-level concepts such as \textit{DATA SUBJECT RIGHT} that is mentioned earlier.  
Level~2, shaded light green (dashed-edge boxes), and level~3, shaded gray (dotted-edge boxes), capture the different specializations. For instance, \textit{DATA SUBJECT RIGHT} has been refined into eight specializations, one for each of the individuals' rights,  such as \textit{ACCESS}, i.e., the right to access personal data. 
For more details, we refer the reader to the original paper~\cite{Amaral:21} \rev{as well as the glossary in the annex~\cite{annex}.} 

\revv{While the model was originally applied in the context of 
privacy
	policies, we note that privacy policies are legally significant
	artifacts that are expected to reflect the actual collection,
	processing, sharing, and management of personal data 
	performed
	by mobile applications. 
  Therefore, the concepts represented in
  the model are not limited to privacy policies themselves but 
  are
  also applicable to the broader analysis of GDPR-related 
  privacy
  concerns in mobile apps. In this study, we map the primary
  studies to the concepts defined in the original model to 
  obtain a
  structured overview of the GDPR concerns investigated in the
  literature.}

\rev{In the remainder of the paper, we will use the term \textit{concept} to refer to the elements of the conceptual model, and the term \textit{concern} to refer to concerns in the GDPR in a broader sense. 
}

\section{State of the art}
\label{sec:related}

Despite the significant impact of GDPR on mobile app 
development and the risks for non-compliance, 
there is no study to date that comprehensively reviews the research landscape on how GDPR-relevant privacy concerns are addressed in mobile apps. 
Below, we discuss secondary studies that survey relevant topics.  

Privacy concerns have been broadly discussed in the SE 
literature. 
Sangaroonsilp and Dam~\cite{Sangaroonsilp2025}  conducted 
a systematic mapping study on privacy-related trends in the 
SE literature between 2016 and 2023, focusing on 13 
predefined privacy threats, such as access control and user 
tracking. 
The authors analyzed 187 primary studies, and 
reported that  
only 187 studies out of the 1540 totally collected (12.14\%) directly addressed privacy.  
The authors further indicated that, despite the central role of regulations, such as the GDPR, the majority of primary studies did not explicitly reference any regulations or standards. 
\revv{
	Compared to this study, our review adopts
	a focused and detailed GDPR-centered analysis aimed to
	characterize the literature as well as identify which 
	GDPR 
	concerns have been extensively investigated and which 
	remain under-explored in the context of mobile applications, 
	potentially leading to non-compliance risks. 
	The few overlapping primary studies (only three) are 
	included in the mapping study among a broader set of 
	software engineering 
	publications, leaving out details related to the GDPR specific 
	concerns. The substantially different objectives and analysis 
	criteria are evidenced in the data extraction fields among 
	the two reviews. } 

\rev{Showail et al.~\cite{Showail2025} examined Saudi Arabia’s Personal Data Protection Law (PDPL) and its application on 66 mobile apps from the Google Play Store and the Apple App Store, using a compliance rubric aligned with PDPL requirements.
The analysis covered five concerns, namely user data rights, data protection principles, privacy policies, breach notifications, and cross-border data transfer practices. 
The results pointed out a compliance gap between the 
regulatory requirements and actual mobile app practices. 
}

More related to our work, Ebrahimi et al.~\cite{Ebrahimi2021} analyzed 59 papers related to privacy in mobile apps. They focused on the practice of obtaining explicit consent prior to accessing sensitive private information, in exchange for offering a more personalized user experience. The authors argue that such privacy-invading practices have led to major privacy concerns among app users. The authors also showed that existing literature had focused mainly on detecting data leaks, with little to no attention being given to the user's perspective. 

In a similar vein, Alloghani et al.~\cite{Alloghani2020} focused on identifying security and privacy issues in mobile devices and systems alongside existing methods for detecting and preventing such issues. The authors emphasized the importance of utilizing preventive, detective, and responsive methods for ensuring security and privacy in mobile applications. Such methods require the involvement of both mobile vendors and service providers. 

Along similar lines, Shrivastava et al.~\cite{Shrivastava2020} analyzed privacy issues concerning permission requests in Android apps. The authors studied 110 research papers, identifying possible severe security implications of the Android permission protocol and further describing several limitations in permission checks. 

More focused on the GDPR, Negri-Ribalta et 
al.~\cite{Negri2024} surveyed the impact of GDPR's on 
regulatory requirements pertinent to  data protection and how 
they are handled from an RE perspective. Their study involved 
90 papers published between 2016 (i.e., right about when the 
regulation was voted) and 2022 (four years after GDPR was 
actually enforced). 

\revv{Among the above studies, the last one is the most 
relevant to ours, as it is also GDPR-centric, albeit it provides a 
generic view of the GDPR provisions. For the analysis, the 
authors created a taxonomy of GDPR privacy concerns 
according to an ad-hoc manual analysis of the regulation. In 
comparison, our work is driven by a comprehensive and 
detailed conceptualization of privacy-relevant concerns in 
GDPR. Mapping existing studies in the literature against this 
conceptual model provides guidance for future comparisons. }

Fernández-Alemán et al.~\cite{Fernandez-Aleman2013} studied security and privacy concerns in electronic health records (EHR) systems. Specifically, the authors analyzed 49 papers, of which only 26 use standards or regulations regarding privacy and security of EHR data. The authors reported the Health Insurance Portability and Accountability Act (HIPAA)~\cite{HIPAA} and the European Data Protection Directive 95/46/EC~\cite{EU95} as the most widely used regulations (we note that the paper was published prior to the GDPR). The authors concluded that, to implement secure EHR systems, extensive harmonization is first necessary to resolve discrepancies between the set of applicable regulatory texts and standards. 

Similarly, Iwaya et al.~\cite{Iwaya2020} analyzed 365 papers 
to study the state-of-the-art on security and privacy in mobile 
health (mHealth) and ubiquitous health (uHealth) systems. 
Results suggested that existing work primarily focused on 
certain aspects that were not necessarily critical to the 
organizations employing mHealth and uHealth systems, 
leaving out aspects such as data governance, security and 
privacy policies, or program management. 

Martínez-Pérez et al.~\cite{Martinez-Perez2014} surveyed 
privacy and security in mHealth apps. Among the 169 analyzed 
papers,   
the authors highlighted the lack of awareness about security and privacy laws and further presented some recommendations to serve as a quick guide for designers, developers, and researchers. Such recommendations provide a more detailed description of the legal text, facilitating compliance to the current security and privacy standards. 

The above studies broadly mention privacy regulations such 
as HIPAA, 
but our work focuses specifically on GDPR and systematically 
analyzes the mobile-app literature using a GDPR-centered 
classification framework based on fine-grained GDPR privacy 
concepts.

\textit{Privacy-by-Design (PbD)} is a central principle in the 
GDPR that has been as well widely studied in SE. 
PbD has largely impacted the drafting of various privacy 
concerns in the GDPR , e.g., the closely related concerns 
data-protection-by-design and by-default (as discussed in art. 
25). We thus review the related work focusing on PbD. 

\rev{To this end, de Chaves et al.~\cite{deChaves2023} reported on the level of understanding of PbD principles and goals in SE during the early years of the GDPR (2018--2022).
Results showed that no particular principle was emphasized in the
survey. The majority of the papers addressed compliance at large with
limited or  no references to a specific privacy principle. 
The contributions of these papers were mostly related to supporting business processes.
}

Similarly, Andrade et al.~\cite{Andrade2023} investigated the 
misuse of personal data and what this entails in terms of 
individuals' privacy and software product's quality. The 
authors analyzed 75 papers to understand how PbD 
principles had been applied in the SE practices.
The results highlighted the lack of specific methodology and tools for translating these principles into practical activities throughout the software development lifecycle. 
This concern has become even more relevant now that PbD \rev{has largely influenced} regulations such as GDPR. 

\rev{In a similar work, Saltarella et al.~\cite{Saltarella2021} investigated the best practices of GDPR-compliant software design and development accounting for the PbD and privacy-by-default paradigms.
The authors focused on two main categories namely, data-oriented strategies and business-oriented strategies and eight sub-categories (four for each).
Results showed that most papers discussed business-oriented strategies, often lacking methodological discussions. 
Data-oriented strategies, on the other hand, largely focused on anonymization, pseudonymization, and encryption (rather than data minimization).}

In a similar vein, Semantha et al.~\cite{Semantha2020} analyzed PbD in the healthcare domain. 
Their work focused on producing recommendations to reduce personal data breaches. 
The authors targeted contemporary frameworks regularly 
applied for safeguarding data privacy; specifically, they 
examined seven PbD frameworks and identified key limitations 
in these frameworks that would allow for potential data 
breaches, 
e.g., inefficiencies in data managing jeopardizing the reliability of personal data. 
Finally, the authors advocated for refining and improving these 
frameworks as a way to reduce the rate of data breaches 
particularly in the healthcare domain. 

Finally, Morales-Trujillo et al.~\cite{Morales-Trujillo2019} analyzed \textit{privacy by design (PbD)} in software systems. 
Their work aimed to identify relevant literature collecting PbD goals (e.g., data protection mechanisms) in software development and determining the extent to which PbD was present in the current software development practices. 
The authors analyzed 49 papers and concluded that the primary studies mainly focus on data minimization. 
The authors further suggested that PbD was still an underdeveloped concept in the SE literature and acknowledged the necessity for a framework to support the development of privacy-aware systems. 

\revv{While principles are important in legal procedure and 
literature, software compliance is rather ensured by extracting 
requirements from the directly applicable parts of regulations. 
However, motivated by the research efforts described under 
this category, our study complements this model, we refer to 
the set of principles defined in Art. 5 in the GDPR, such as data 
minimization, purpose limitation, or fairness, transparency, and 
accountability.}

In summary, Table~\ref{tab:related} compares the 
aforementioned secondary studies against ours. We note that 
only two studies have explicitly considered \rev{the GDPR, 
including 
Negri-Ribalta et al.'s work~\cite{Negri2024} on general privacy concerns under the GDPR and Saltarella et al.'s survey concerning PbD~\cite{Saltarella2021}. }
The majority of the remaining studies have no explicit legal 
ground in their work.  
In contrast, in this SLR we present a comprehensive overview of the existing research landscape in SE, focusing on privacy in mobile apps.
Our SLR further projects the current research onto a \textit{reference model} of GDPR privacy concepts.

\begin{table*}
\small
\centering
\caption{Analysis of the State of the Art}
\label{tab:related}
  \centering
   \begin{tabularx}{0.98\linewidth}{@{} p{0.25\linewidth} @{\hskip 0.5em} p{0.1\linewidth} @{\hskip 0.5em} p{0.1\linewidth} @{\hskip 0.5em} p{0.15\linewidth} *{1}{>{\arraybackslash}X}@{}}
\toprule
   & Grounding Legal Basis & Studies & Privacy-related Concepts  & Focus \\
   \midrule
\rev{Sangaroonsilp \& Dam~\cite{Sangaroonsilp2025}} & \rev{N/A} & \rev{187} & \rev{13 privacy threats} & \rev{Privacy trends in the SE literature}\\
\rev{Showail et al.~\cite{Showail2025}}&\rev{PDPL}
&\rev{66}&\rev{5 privacy concerns} & \rev{Saudi Arabia’s Personal Data Protection Law (PDPL) compliance in mobile apps} \\
Ebrahimi et al.~\cite{Ebrahimi2021} & N/A & 59 & N/A & Consent and access to personal information in mobile apps\\
Alloghani et al.~\cite{Alloghani2020} & N/A & 330 & N/A & Security and privacy in mobile devices and systems\\
Shrivastava et al.~\cite{Shrivastava2020} & N/A & 110 & 1 (Consent) & Security and Privacy of Android Permission System \\
Negri-Ribalta et al.~\cite{Negri2024} & The GDPR & 90 & 30 & Requirements related to the GDPR in the literature \\
Fernández-Alemán et al.~\cite{Fernandez-Aleman2013} & N/A & 49 & 5 high-level concerns & Security and privacy in electronic healthcare records \\
Iwaya et al.~\cite{Iwaya2020} & NIST-8062 & 365 & 20 controls from the standard & Security and privacy in mHealth and uHealth systems\\
Martínez-Pérez et al.~\cite{Martinez-Perez2014} & N/A & 169 & 7 high-level concerns & Security and privacy in mHealth apps\\
\rev{de Chaves et al.~\cite{deChaves2023}} & \rev{N/A} & \rev{68} & \rev{7 PbD principles} & \rev{Level of understanding of PbD in the SE literature } \\
Andrade et al.~\cite{Andrade2023} & N/A & 75 & N/A & Privacy by design\\
\rev{Saltarella et al.~\cite{Saltarella2021}} & \rev{The GDPR} & \rev{91} & \rev{7 PbD principles} 
& \rev{Best practices of design and development for the PbD 
and privacy-by-default paradigms} \\
Semantha et al.~\cite{Semantha2020} & N/A & 23$^*$ & 7 PbD principles & Privacy by design in healthcare\\
Morales-Trujillo et al.~\cite{Morales-Trujillo2019} & ISO/IEC 29100 & 49 & 11 ISO/IEC 29100 principles & Privacy by design\\
\midrule
\textbf{This work} & The GDPR & 60 & 56 & GDPR-related Privacy in Mobile Apps  \\
\bottomrule
\end{tabularx}
 \begin{tablenotes}
   \item \it $^*$ The SLR included up to 253 studies but the mapping study on PbD principles concerned 23 studies.
\end{tablenotes}
\end{table*}

\section{Study Design}
\label{sec:studyDesign}

\begin{figure} 
\centering
  \includegraphics[width=0.49\textwidth]{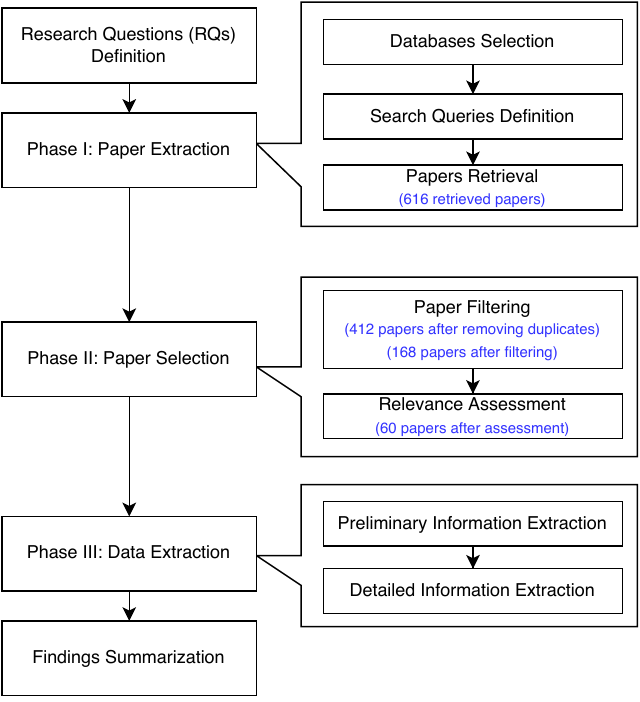}
  \caption{Overview of our Review Process}
  \label{fig:approach}
\end{figure}

We conducted this SLR following the recommended guidelines by Kitchenham and Charters~\cite{KitchenhamCharters2007,Petersen2008,Guidelines2015}. 
Fig.~\ref{fig:approach} shows the different steps that we followed to achieve our findings. 
In particular, our method spanned three phases, namely (i) searching for the literature (i.e., paper extraction), (ii) selecting the papers to analyze (i.e., paper selection), and (iii) carrying out the data extraction. To this end, the first author defined the review protocol, performed the search and selection of relevant papers, and performed the data extraction step. This process was closely supervised by the second and third authors. Findings were thoroughly discussed and validated across several sessions.

\subsection{Research Questions (RQs) Definition}
\label{subsec:rqs}

This study investigates four research questions (RQs):

\sectopic{RQ1. How well does the literature on privacy in 
mobile apps cover \rev{the GDPR privacy concepts?}}
This RQ explores the research landscape on mobile apps to identify the main topics in terms of GDPR privacy  \rev{concerns}. 
Drawing on the \textit{reference model} (presented in Section~\ref{sec:background}), we answer RQ1 by mapping the topics investigated in existing work on mobile apps to the \rev{concepts} in the conceptual model.

\sectopic{RQ2.  What are the main objectives pursued when investigating GDPR privacy  \rev{concerns}?}
RQ2 explores the main objectives of primary studies that investigated GDPR privacy in mobile apps. 
\rev{The intuition behind this RQ is to understand the \emph{GDPR-related research challenges}, which were acknowledged or addressed by existing work.
As a result, RQ2 identifies remaining challenges that are currently under-explored and which can lead to potentially open directions for future research. }

\sectopic{RQ3. What are the different types of research contribution?}
RQ3 explores the different types of research contribution made by the primary studies, distinguishing between \emph{technical} and \emph{non-technical} contributions, and analyzing the app stores and operating systems targeted in the literature. 
It provides a comprehensive view on the research 
contributions and further describes, in the case of technical 
solutions,  the enabling technologies and the way such 
contributions were validated.  

\sectopic{RQ4. Are the research artifacts accompanying studies on GDPR privacy  \rev{concerns} publicly available?}
This RQ sheds light on the publicly available artifacts that can be potentially leveraged in future research. 
RQ4 further reports on the licenses under which these artifacts have been shared. 

\subsection{Phase I: Paper Extraction}
\label{subsec:extraction}

Ensuring the retrieval of a relatively representative sample of the existing literature is essential for conducting an SLR.
This phase involves three steps: (1) selecting the databases or search engines to query, (2) refining the query keywords, and (3) \rev{retrieving the primary studies}.  
Next, we elaborate on each step. 

\subsubsection{Databases Selection }

For the search of papers in the different databases, we adopted common practices for and recommendations from the SLR literature~\cite{Dyba2007,Guidelines2015,Petersen2015}. Specifically, we retrieved primary studies from the following databases: ACM Digital Library\footnote{\url{https://dl.acm.org/}}, IEEE Xplore\footnote{\url{https://ieeexplore.ieee.org/}}, ScienceDirect\footnote{\url{https://www.sciencedirect.com/}}, Scopus\footnote{\url{https://www.scopus.com/}}, and SpringerLink\footnote{\url{https://link.springer.com/}}. 
These sources are typically selected in SLRs and systematic mapping studies~\cite{zhao2021,montgomery2022}.
Additionally, we searched for papers in the following search 
engines or indexing systems: 
DBLP\footnote{\url{https://dblp.org/}}, Google 
Scholar\footnote{\url{https://scholar.google.com/}}, and Web 
of Science\footnote{\url{https://www.webofscience.com/wos/}}.
 
\rev{The rationale behind including these additional sources is mainly to broaden the search results and ensure a complete coverage of the SE literature}. Such indexing systems typically contain preprints or in-press papers that are not yet included in the regular databases mentioned above, and would be otherwise missed. 
We scoped our search against these indexing systems by 
restricting it to the period 2022--2023, \rev{as we assume the 
papers published before 2022 are already covered by the 
comprehensive set of results retrieved from the main 
databases. 
We acknowledge that restricting the scope to this two-year period may introduce a threat to validity, as it can lead to missing potentially relevant studies that not indexed in the primary databases. We elaborate on this validity consideration in Section~\ref{sec:threats}.}

\revv{It is worth noting that the search and data-collection 
process was conducted in 2023,
with the subsequent data analysis completed in early 2024.
Accordingly, we restricted the search period to publications
available up to 2023, ensuring that the review covered the 
most
recent literature accessible at the time of data collection while
maintaining a reproducible search scope. }

\subsubsection{Search Query Definition }

\revv{To reduce the risk of missing relevant studies due to
terminological variations, we iteratively refined our search 
queries by experimenting with several terms targeting papers 
that address \textit{GDPR privacy in mobile apps}. In addition, 
search was performed against the bibliographic fields
supported by the selected indexing systems, including titles,
abstracts, and keywords whenever permitted by the search
interface. }

\rev{Initially, we tried different combinations of relevant keywords like ``data protection'', ``privacy'', and ``mobile applications''. We observed that including some keywords would narrow the search results while others broaden them beyond the scope of our study, yielding largely irrelevant papers. 
We tested several queries across the different databases.} 
As a result, we deemed the following keywords to be the best 
fit for this study:    (``mobile app*'' AND	``data protection''	
AND ``data privacy''). 
We queried all databases using the same query except for ScienceDirect, which does not allow using special characters such as wildcards(\texttt{*}).  
In this case, we used the following combination: (``mobile 
app''	OR	``mobile apps''	OR 	``mobile application''	OR	
``mobile applications'').

Finally, \textit{our final query } is as follows:

\begin{tcolorbox}[colback=black!5!white,colframe=white!50!black,]
\rev{\texttt{``mobile app'' AND ``privacy'' AND ``gdpr''} }
\end{tcolorbox}

\subsubsection{Papers Retrieval } 
\label{subsec:paper_retrieval}

In this step, we queried the various databases using our defined queries. All  queries were applied on the full text of the papers. 
Additionally, we extracted papers while directly integrating the following inclusion criteria:  

\begin{itemize}
    \item[\textbf{I1}] \textit{Time span:} To fully capture primary studies focusing on GDPR, we defined the time span 2016--2023 early in our search process. The GDPR was adopted in 2016 after passing European Parliament and entered into force on May 25, 2018. 
    Using this time span, we targeted research work that investigated GDPR not only after its entry into force, but also its final public version shortly before. 

    \item[\textbf{I2}] \textit{Language:} We exclusively considered primary studies written in English. 

    \item[\textbf{I3}] \textit{Domains and paper types:} We 
    scoped our search to subject area of \textit{Computer 
    Science} OR \textit{Engineering}, and we further limited the 
    search to \textit{Conference papers} OR \textit{Journal 
    Articles}.  
\end{itemize}

\begin{table}
\centering
\caption{Number of Papers Initially Retrieved per Database 
using the query \texttt{``mobile app'' AND ``privacy'' AND 
``gdpr''} and Remaining Papers after Filtering and Relevance 
Assessment. }
\label{tab:final_queries}

  \begin{tabularx}{0.48\textwidth}{@{} p{0.2\textwidth} @{\hskip 0.5em} *{3}{>{\centering\arraybackslash}X}@{}}
     \toprule
    Database & Retrieval Results &  \rev{Filtering Results} & \rev{Relevant Papers} \\
    \midrule
    ACM Digital Library	&	181		&	61	&	23	\\
    IEEE Xplore	        &   70		&	18	&	13	\\
    ScienceDirect	    &	129		&	43	&	11	\\
    Scopus	            &	26		&	5	&	5	\\
    SpringerLink	    &	78		&	41	&	8	\\
    DBLP                &	15	    &	0	&	0	\\
    Google Scholar	    &	100	    &	0	&	0	\\
    Web of Science	    &	17		&	0	&	0	\\
    \midrule
    Summary	            &	616     &	168	&	60	\\
    \bottomrule
 \end{tabularx}

\end{table}

 \rev{We applied the above criteria as filters across the selected
   databases. However, slight adaptations were necessary to
   accommodate the available options in each platform. For instance,
   criteria \textit{I1--I3} were fully applicable in ScienceDirect, Scopus, and SpringerLink. IEEE Xplore and the ACM Digital Library do not allow for explicit domain specification. This limitation is nonetheless acceptable since both platforms are focused on computer science and related scientific fields, and hence any irrelevant papers were subsequently excluded during the assessment step. 
 	In ACM Digital Library, journal (articles) and conference proceedings must be selected separately, requiring manual merging of the resulting lists. In DBLP, filters must be split by paper type and publication year, with results to be aggregated afterwards. Web of Science lacks options for specifying the domain, publication period, or paper type. Similarly, GoogleScholar does not support filtering by paper type or research domain. 
 }
 
The number of retrieved papers from each source are listed in \rev{the
  second column of} Table~\ref{tab:final_queries}. \revn{We 
  note that
  the results presented in this study reflect the state of the art 
  as of December 2023.}
Querying the regular databases resulted in a collection of 484 papers, distributed across the different databases as follows: 181 from ACM Digital Library, 70 from IEEE Xplore, 129 from ScienceDirect, 26 from Scopus, and 78 from SpringerLink. 
As noted earlier, we also queried indexing systems for additional potentially missing papers. 
Consequently, we retrieved 132 additional papers, including 100 papers from Google Scholar, 17 papers from the Web of Science, and 15 papers from DBLP. 
As for Google Scholar, we considered only the first 10 pages of results\rev{; we checked the results in the following pages and confirmed that they were not relevant to the objectives of our study. }
\rev{
\revv{Following best practices, we performed snowballing and further applied 
the
search string to dedicated venues. This process did not 
lead to the identification of any new papers that had not already been identified 
during our
paper retrieval step.
}
} 
Our paper extraction phase resulted in a total of \textit{616 papers}, which were then passed on to the next phase for further analysis.

\subsection{Phase II: Paper Selection}
\label{subsec:selection}

In this phase, we selected papers that are relevant for our study, following three steps including: (1) removing duplicate papers and cleaning the results, (2) excluding papers according to certain criteria, and (3) identifying the papers within the scope of our study. 
We explain these steps below.  

\subsubsection{Paper filtering }

\rev{The first step of the paper filtering consisted in removing the duplicates from the entire collection of 616 papers.}
To identify duplicates, we defined a case-insensitive 
\textit{Excel macro} comparing the titles of the papers and 
determining possible duplicates. 
The output of this macro indicated which paper metadata 
details (particularly the title) were actually duplicates. This 
output was then manually vetted by the first author of this 
paper. 
As a result, we filtered out 204 papers which were deemed genuine duplicates, thereby reducing our collection to 412 papers.

To further refine the retrieved primary studies, we applied two \textit{exclusion criteria}, described below.

\begin{itemize}
    \item[\textbf{E1}] \textit{Paper length:} We defined a 
    minimum paper size of six pages (including references). The 
    rationale behind this threshold was to 
    discard short papers such as position, vision, demo, new-idea, early-research-achievements papers with only preliminary or partial results, as commonly done in relevant literature~\cite{zhao2021,montgomery2022}.

    \item[\textbf{E2}] \textit{Publication venues:} 
    Following the guidelines of Kitchenham and Brereton~\cite{KitchenhamBrereton2013}, we filtered out papers that were not 
    relevant for our study. Specifically, we excluded papers 
    published in journals or conferences clearly unrelated to SE 
    and privacy. In this way, we removed 
    papers focusing on the social aspects of privacy/data 
    protection or investigating specialized topics. In case of 
    uncertainty (e.g., when a paper title appeared relevant), we 
    retained the venue and related paper(s) for further analysis 
    (see below). 
    Examples of excluded venues are the \textit{International Journal of Medical Informatics}, \textit{Blockchain: Research and Applications}, and the \textit{EAI International Conference on Smart Objects and Technologies for Social Good}. 
    
\end{itemize}

\rev{At the end of the paper filtering, we were left with 168 papers; the details of their sources are provided in the third column of Table~\ref{tab:final_queries}.}
The venues (36 journals and 24 conferences) and the 
respective number of papers (105 journal papers and 63 
conference papers) are summarized in column N of 
Tables~\ref{tab:considered_journals}~and~\ref{tab:paper_venues}.
\rev{These papers were then processed through the relevance assessment step.}

\subsubsection{Relevance Assessment:}\label{sec:relev-assessm}
We analyzed the remaining 168 papers and identified those relevant for our SLR. 
The first three authors collectively participated in this process. 
Driven by the motivation of our SLR, the relevance of the papers was assessed according to their main focus being privacy. 

\rev{Therefore, we considered a paper relevant if:} 
\begin{itemize}
\item[\textbf{I4}] \rev{The paper \textit{clearly discussed 
GDPR-relevant privacy concerns and how these were handled 
in mobile apps.} 
For instance, a paper that discusses users practices in mobile apps, such as subscription cancellation and asking for refunds, or users providing inaccurate information were considered as not relevant to our SLR. 
Similarly, papers discussing dark patterns that attempt to manipulate users' behaviors are not relevant to our SLR. 
While these papers could be relevant to privacy in a broader 
sense, they are less informative about how the handling and 
processing of personal data were addressed in mobile apps.}
\end{itemize}

\rev{The relevance assessment step} spanned three iterations, described below. 

In the first iteration, the first author read through all abstracts 
and labeled the respective papers according to three 
categories, namely ``relevant'' when the paper was relevant to 
our SLR, ``not relevant'' when it was not relevant, and 
``possibly relevant'' when the author was in doubt or the 
abstract was not detailed enough for a conclusive decision.  
As a result, 
52 papers were labeled as ``relevant'', 94 as ``not relevant'', and 22 as ``possibly relevant''. 

The second iteration aimed to confirm this labeling, focusing on the 52+94=146 papers deemed as ``relevant'' or ``not relevant''. Specifically, we randomly split the papers into two equal subsets, of 73 papers each. Subsequently, the second and third authors independently cross-checked all papers in these two subsets. This validation process confirmed the decisions made in the first iteration. All disagreements were thoroughly discussed by the three authors.   

In the last iteration, the first author examined the remaining 22 
``possibly relevant'' papers by screening the full paper and 
carefully considering the relevance of its content to the 
objectives of our SLR. As a result, eight papers were deemed 
``relevant'', and the remaining 14 were deemed ``not 
relevant''. 
The second and third authors confirmed the decisions on the 
papers in this iteration as well.

The output of this step is the final set of 60 papers (30 journal papers and \rev{30} conference papers), as indicated in column N$^*$ of Tables~\ref{tab:considered_journals}~and~\ref{tab:paper_venues}), corresponding to 35.7\% of the 168 analyzed papers, on which we conducted our data extraction phase.

\begin{table}[t]
    \caption{Overview of journal articles, assessed (column N) and deemed relevant (column N$^*$) in this SLR.}
\label{tab:considered_journals}

\footnotesize
\centering
 \begin{threeparttable}[t] 
   \begin{tabularx}
   {0.98\linewidth}{@{} p{0.08\linewidth} @{\hskip 0.2em} p{0.79\linewidth} @{\hskip 0.2em} p{0.07\linewidth} @{\hskip 0.2em} p{0.05\linewidth} 
   }
     \toprule
    ID    &   Journal Name & N & N$^*$  \\ 
    \midrule
    J1 & Journal of Computer Science and Technology & 1 & 0 \\ 
    J2    & Journal of Engineering and Applied Science & 1 & 0 \\
    J3    & Mobile Networks and Applications & 1 & 0 
    \\
    J4 & Internet Interventions & 1 & 1\\
    J5    & Annals of Telecommunications & 1 & 0 
    \\
    J6   & Information and Software Technology & 3 & 1\\
    J7    & Wireless Networks & 1 & 1 
    \\
    J8 & Computers \& Security & 12 & 4\\
    J9  & Soft Computing & 1 & 0 
    \\
    J10 & Journal of Systems and Software & 3 & 1\\
    J11   & Complex \& Intelligent Systems & 1 & 0 
    \\
    J12    & Frontiers of Computer Science & 2 & 1 \\ 
    J13 & Journal of Information Security  and Applications & 3 & 0  
    \\
    J14   & International Journal of Information  Technology & 1 & 0 \\
    J15   & Human-Centric Intelligent Systems & 1 & 0 
    \\ J16 & Computer Law \& Security Review & 16 & 4\\
    J17  & AI and Ethics & 2 & 0 \\
    J18   & Software and Systems Modeling & 1 & 1 \\
    J19 & Intelligent Systems with Applications & 1 & 0 \\
    J20   & Computing & 2 & 1 \\
    J21 & Information Processing \& Management  & 1 & 0 \\
    J22  & Empirical Software Engineering & 5 & 2 \\
     J23 & Computer Fraud \& Security & 1 & 0 \\
    J24  & Artificial Intelligence and Law & 1 & 0 \\
     J25 & Information Systems & 1 & 0 \\
    J26  &  International Cybersecurity Law Review & 1 & 0  \\
     J27   & AI \& Society & 8 & 1 
\\ J28  & International Journal of Information Security & 4 & 1 \\
     J29 & Data Science and Management & 1 & 0 \\
    J30  & Ethics and Information Technology & 3 & 0 \\
    J31   &Frontiers of Information Technology  \& Electronic \par Engineering & 1 & 0
    \\
    J32 & ACM Transactions on Computer-Human \par  Interaction & 10 & 7\\
    J33  & Personal and Ubiquitous Computing & 3 & 1 
    \\
    J34 & IEEE Transactions on Software Engineering & 0 & 2\\
    J35 & ACM Transactions on Software Engineering \par Methodology & 2 & 1 \\
    J36 & Proceedings of the ACM on  Computer-Human \par  Interaction & 8 & 0 \\
    \midrule
    \multicolumn{2}{c}{Total} & 105 & 30 \\

    \bottomrule
 \end{tabularx}
  \end{threeparttable}

\end{table}

\begin{table}
    \caption{Overview of conference papers,  assessed (column N) and deemed relevant (column N$^*$) in this SLR.}
     \label{tab:paper_venues}
      \footnotesize
\begin{threeparttable}[t]
\centering
    \begin{tabularx}{\linewidth}
   {@{} p{0.07\linewidth} @{\hskip 0.2em} p{0.8\linewidth} @{\hskip 0.2em} p{0.05\linewidth} @{\hskip 0.2em} p{0.05\linewidth}
   }
    \toprule
    ID    &   Venue & N & N$^*$ \\
    \midrule   
    V1 & International Conference on Mining Software \par Repositories &1&0 
    \\
    V2    & IEEE International Conference on Software \par  Analysis, Evolution and Reengineering & 1 & 1 \\
    V3 & International Conference on Availability, \par Reliability and Security & 3 & 1 \\
    V4 & Annual ACM Symposium on User Interface \par Software and Technology &1&0\\
    V5 & International Conference on Computer Systems \par and Technologies &1&0 \\
    V6   &   Annual Computer Security Applications \par Conference & 1 & 1\\
    V7    & The Web Conference &1&1
    \\
    V8 & ACM Web Conference & 4 & 1 \\
    V9  & Annual International Conference on Mobile \par Computing and Networking & 1 & 0
    \\
    V10 & Annual IEEE/IFIP International Conference \par on Dependable Systems and Network & 1 & 1\\
    V11  & USENIX Security Symposium & 2 & 2
    \\
    V12 & IEEE Symposium on Security and Privacy  &1&0\\
    V13  &  Annual Computers, Software, \par and Applications Conference & 1 & 0 \\
    V14   & ACM SIGSAC Conference on Computer and \par Communications Security & 4 & 2\\ 
    V15   & ACM Conference on Security and Privacy in \par  Wireless and Mobile Networks & 1 & 1 
    \\ 
    V16 & IEEE/ACM International Conference on \par Automated Software Engineering & 6 & 5\\
    V17  & CHI Conference on Human Factors in \par Computing Systems & 10 & 1
    \\ 
    V18 & IEEE International Requirements Engineering \par Conference & 3 & 3 \\
    V19  & ACM SIGSOFT International Symposium on \par Software Testing and Analysis &1&0
    \\
    V20 & IEEE European Symposium on Security \par and Privacy Workshops & 4 & 2\\
    V21  & ACM/IEEE International Conference on \par Software Engineering & 6 & 6
    \\ 
    V22 & IEEE Annual Computer Software and \par Applications Conference & 1 & 1\\
    V23 & IEEE International Conference on Trust, Security \par  and Privacy in Computing and Communications  &1&0 \\
     V24   &   International Conference on Mobile and \par Ubiquitous Multimedia &1&0
    \\
    V25 & Workshop on Privacy in the Electronic Society & 4 & 1 \\
    \midrule
    \multicolumn{2}{c}{Total} & 63 & 30 \\ 
   \bottomrule
 \end{tabularx}
   \end{threeparttable}

\end{table}

\subsection{Phase III: Data Extraction}
\label{subsec:dataextraction}
Our data extraction phase targeted certain \textit{relevant 
information}, listed in Table~\ref{tab:extraction}. 
In addition to the metadata about the paper (prefixed by C0 in the table), e.g., its title and the authors name, we extracted the following information: 
the GDPR concepts and principles covered in the primary studies (C1), the traced personal data (C2), the targeted app
store(s) and operating system(s) (C3), the goals and contributions of the primary studies (C4), the applied data analysis methods (C5), the proposed solutions and their enabling technologies (C6), the data used (C7), and whether any artifact was made publicly
available (C8).
We extracted this information by thoroughly reviewing the 60 papers marked as relevant in the previous phase. 
\rev{Our data extraction followed an integrated coding approach~\cite{cruzes2011}, combining both inductive and deductive analysis. We started our analysis with a predefined set of codes derived from the concepts outlined in our reference model. However, as not all primary studies explicitly stated these concepts explicitly, we also based our extraction on interpretive  analysis to identify implicit evidence pertinent to the concepts. Consequently, our findings draw both on the verbatim content of the studies and on our interpretation in relation to the reference model. This hybrid method enabled a more comprehensive and grounded analysis of the data.  }

\rev{The fields prefixed with C1.* are intended for addressing 
RQ1,
  while those prefixed with 
C2.*, C3.*, and C4.* are intended for addressing RQ2, with C3 providing a focused view on the possible app market restrictions scoped in the primary studies.
Fields prefixed with C4.*, C5.*, C6.* are designed to answer RQ3,
while  those prefixed with C7.* and C8.* are intended to answer RQ4.}

\rev{Regarding C1 fields, our coding activity involved labeling the primary studies with the relevant set of GDPR principles listed in Art.~5 (C1.1), the set of concepts from our reference model (C1.2), and any other GDPR-relevant concern  not captured by the first two fields (C1.3). 
The motivation for including C1.3 is to make our findings more informative to the research community. A single primary study can obviously contain mentions of multiple concepts, principles, other concerns, or a combination thereof. 
The remaining fields capture information that are not strictly dependent on GDPR principles or concepts.}

The coding was primarily performed by the first author. 
	\revv{To increase confidence in the study-selection and 
	data-extraction
		processes, two authors independently validated two 
		disjoint 
		subsets
		covering 10\% of the primary studies each (20\% in total). 
		The
		objective of this validation was to assess consistency in 
		the
		application of the selection and extraction criteria rather 
		than
		to obtain a statistically representative sample. Any
		disagreements were discussed among the authors and 
		resolved
		through consensus.} 
We then measured the agreement between the concepts identified by the first author and those identified by the second and third authors using the Krippendorff's alpha coefficient~\cite{krippendorff1970}. 
We obtained an agreement value of 0.94, indicating 
\textit{high agreement} among the coders.

\begin{table*}
\centering
\caption{Data Extraction Fields }
\label{tab:extraction}
\begin{threeparttable}
\centering
\begin{tabularx} 
{0.98\linewidth}{@{} @{\hskip 0.4em}  p{0.07\linewidth} @{\hskip 0.2em}  p{0.2\linewidth} @{\hskip 0.2em} *{1}{>{\arraybackslash}X}@{}}
\toprule
ID & Label   & Description  \\
    \midrule
    C0.1 & Paper ID  & Each paper is assigned a unique identifier. \\
    C0.2 & Paper Type &  Whether the paper is a conference paper or a journal article.  \\
    C0.3 & Title   &   The title of the paper. \\
    C0.4 & Authors   &  The authors names. \\
    C0.5 & Venue  & The name of the journal or conference where the paper is published. \\
    C0.6 & Publication Year & The year when the paper is published. \\
    C0.7 & Pages	& The number of pages in the paper. \\ 
    C0.8 & Publisher	& The name of the publisher. \\
    C1.1 & GDPR Principles	& Key GDPR principles \rev{(according to GDPR Art. 5)} mentioned in the paper. \\
    C1.2 & Privacy Concepts & Key GDPR privacy concepts mentioned in the paper according to our \textit{reference model}\tnote{§}. \\
	\rev{C1.3} & \rev{Other Concerns} & \rev{Other GDPR-related concerns mentioned in the paper.} \\
    C2.1 & Personal Data	& The list of personal data that is mentioned and which the mobile apps tracks. \\
    C3.1 & OS & The operating system (OS) investigated by the paper. \\
    C3.2 & App Store &	The app store investigated by the paper. \\
    C4.1 & Main Goal & The main goal of the paper.\\
    C4.2 & Contributions & The key contributions of the paper.\\
    C5.1 & Study Type & The study type conducted in the paper. \\ 
    C5.2 & Research Approach & Whether the paper conducts qualitative or quantitative analysis, or a  mix of both. \\ 
    C5.2.1 & Participants & The number of participants in the 
    qualitative study. \\ 
    C6.1 & Solution Type & Whether the solution is purely manual, semi-automated, or automated. \\
    C6.2 & Enabling Technology & The enabling technologies 
    used to build the proposed solution in the paper. \\
    C7.1 & Data Source	& The source of the data which is used in the paper.  \\
    C7.2 & Domain	& The domain of the data used in the paper. \\
    C7.3 & Dataset Size   & The size of the dataset or corpus used in the paper. \\
    C7.4 &  Dataset Categories	& The categories of the data and mobile apps collected and/or used in the paper.\\
    C8.1 & Availability & Any links provided with the paper with shared material. \\
\bottomrule
\end{tabularx}
\end{threeparttable}
\end{table*}

Our data extraction process involved two steps, described below.

\subsubsection{Preliminary Information Extraction} 

In this step, we primarily reviewed the abstract and the ``introduction'' section (together with the ``background'' and ``related work/state of the art'' sections if needed to better understand the paper). 
The purpose of this step is to extract generic \textit{relevant information} related to the main focus of the primary study. 
Concretely, we extracted the main goal, the contributions, the research questions, 
and information on online availability of resources, if applicable. 
We also identified the covered GDPR principles and privacy 
concepts. 
When such details, e.g., concerning contributions or research questions, were not clearly stated in the analyzed papers, we labeled the respective information as ``not stated''. 

\subsubsection{Detailed Information Extraction} 

In the last step, we analyzed the remaining sections of the papers. 
We reviewed the details related to the proposed approaches or automated solutions, experimental design, implementation, evaluation, and results. 
From these sections, we extracted additional \textit{relevant information}. 
In the cases where a type of \textit{relevant information} was 
not applicable or not clearly found in the analyzed paper, we 
labeled it as ``not stated''. 

The extracted \textit{relevant information} from all relevant papers are the basis for our findings and for answering our RQs next.

\subsection{Data Availability}
We make the raw data of our study available to reviewers 
\rev{in an annex~\cite{annex}}. 
We plan to make it publicly available upon acceptance of the 
paper. 

\section{Findings}
\label{sec:findings}

In  this section, we answer the RQs outlined in 
Section~\ref{sec:studyDesign}.

\subsection{RQ1. How well does the literature on privacy in mobile apps cover the GDPR privacy \rev{concepts}?}
\label{sec:rq1}

To answer RQ1, we mapped the privacy \rev{concerns} highlighted in the primary studies to the \rev{concepts} in our \textit{reference model} \rev{and the list of principles from Art. 5 of GDPR} (see Section~\ref{sec:background}).

According to our \rev{observations}, \rev{48} out of 60 papers (\rev{80\%}) mention one or more GDPR privacy concepts,  \rev{19} papers ($\approx$32\%) mention one or more principles (from \rev{Art. 5}), and 
\rev{18 papers (30\%) mention other privacy concerns, besides the concepts and principles. 
Other concerns are mostly related to privacy or security in a broader
sense than what is in the GDPR, including privacy by design, 
data anonymization, 
data transfer to third parties\footnote{Note that ``Third Party'' in Table~\ref{tab:coverage} indicate collecting data from third-parties (not transferring to them).}, and notification in case of data breach.
Finally, five papers ($\approx$8.3\%) lack mentions of any GDPR principle or concept, but they refer to privacy at large, data protection at large, privacy protection, privacy by design, security, and security by design. These papers are typically about privacy risks rather than rights and obligations under GDPR.}

51 out of 60 papers ($\approx$85\%) have been published after 2020, with half of them published in 2023.
These percentages indicate the increase of interest in investigating the GDPR privacy concepts in various SE contexts, as we will show later in this section. 
\rev{Table~\ref{tab:coverage} presents the mapping between the GDPR concepts and the analyzed papers.
Following our reference model, we organize the concepts in the table
into three layers (L1, L2, and L3), with L3 being a specialization of
L2, which is in turn a specialization of L1. We use separators between
the different L1-concepts (grouped with their L2 and L3
specializations). Following this, we count the number of primary
studies as follows: whenever a concept is identified at levels L2 or
L3, we also count the paper in the parent concept (L1 or L2, respectively), e.g., a paper discussing \textit{WITHDRAW CONSENT} (L2) is by default also discussing \textit{DATA SUBJECT RIGHT} (L1), since the former is a specialization of the latter.
} 
Note that the total number of papers exceeds \rev{48}, since 
the same paper could discuss multiple \rev{concepts}.

\begin{table*}
\centering
\small
\caption{Mentions of \rev{GDPR Concepts} in the Literature (RQ1) }
\label{tab:coverage}
\begin{threeparttable}
\centering
\begin{tabularx} 
{0.98\linewidth}{@{} @{\hskip 0.4em}  p{0.3\linewidth} @{\hskip 0.2em} p{0.1\linewidth} @{\hskip 0.2em} p{0.05\linewidth} *{1}{>{\arraybackslash}X}@{}}
\toprule
\rev{Concept}       & Level\tnote{§} & Total & Papers \\
\midrule
Recipients              & L1 & 26 &  \citeS{S2,S3,S7,S8,S12,S13,S14,S17,S23,S28,S29,S31,S35,S38,S39,S40,S41,S44,S46,S47,S48,S51,S52,S53,S54,S59}  \\
\midrule   
Children                & L1 &  7 & \citeS{S2,S9,S10,S24,S32,S37,S40} \\
\midrule
PD Security             & L1 &  3 & \citeS{S2,S11,S40} \\ 
\midrule
PD Time Stored          & L1 &  9 & \citeS{S2,S7,S11,S14,S23,S30,S37,S40,S53} \\
\midrule
Auto Decision Making    & L1 &  3 & \citeS{S12,S14,S44} \\
\midrule
Processing Purposes     & L1 &  9 & 
\citeS{S2,S7,S11,S14,S17,S23,S30,S37,S43} \\ 
\midrule
DPO                     & L1 &  1 & \citeS{S30} \\
Contact                 & L2 &  1 & \citeS{S30} \\
\midrule
Controller              & L1 &  2 & \citeS{S30,S40} \\
Contact                 & L2 &  2 & \citeS{S30,S40} \\
\midrule
Controller Representative             & L1 &  0 & \\
\midrule
Data Subject Right      & L1 &  9 & 
\citeS{S2,S10,S14,S30,S37,S40,S44,S53,S56} \\ 
Withdraw Consent        & L2 &  3 & \citeS{S44,S53,S56} \\
\midrule
PD Category             & L1 & 13 & \citeS{S11,S14,S15,S41,S48,S51,S52,S54,S55,S56,S57,S58,S60} \\
Special                 & L2 &  3 & \citeS{S14,S41,S58} \\ 
Type                    & L2 &  1 & \citeS{S54} \\
\midrule
PD Origin               & L1 & 32 &  \citeS{S2,S3,S7,S11,S12,S13,S14,S15,S17,S23,S28,S29,S30,S31,S34,S35,S37,S38,S39,S40,S41,S43,S44,S48,S51,S52,S53,S54,S56,S58,S59,S60}   \\
Direct                  & L2 & 31 &  
\citeS{S2,S3,S7,S11,S12,S13,S14,S15,S17,S23,S28,S29,S30,S31,S34,S35,S37,S38,S39,S40,S41,S43,S48,S51,S52,S53,S54,S56,S58,S59,S60}
   \\ 
Indirect				& L2 &  3 & \citeS{S44,S53,S54} 		\\
Third Party             & L3 &  2 & \citeS{S53,S54} \\
Cookie                  & L3 &  1 & \citeS{S44} \\
\midrule
Transfer outside Europe & L1 &  1 & \citeS{S53} \\
Safeguards              & L2 &  1 & \citeS{S53} \\
\midrule
Legal Basis             & L1 & 21 & \citeS{S1,S5,S7,S11,S14,S22,S32,S34,S38,S39,S41,S42,S43,S44,S45,S48,S49,S50,S53,S56,S58}  \\
Consent                 & L2 & 21 & \citeS{S1,S5,S7,S11,S14,S22,S32,S34,S38,S39,S41,S42,S43,S44,S45,S48,S49,S50,S53,S56,S58} \\
Legitimate Interest     & L2 &  1 & \citeS{S49}  \\
Vital Interest          & L2 &  1 & \citeS{S49}  \\
Contract                & L2 &  1 & \citeS{S49}  \\
Public Function         & L2 &  1 & \citeS{S49}  \\
Legal Obligation        & L2 &  1 & \citeS{S49}  \\
\midrule
PD Provision Obliged             & L1 &  0 & \\   
\bottomrule
\end{tabularx}
\begin{tablenotes}
 \item[§] Level corresponds to the hierarchy in the 
 \textit{reference model} as shown in Fig.~\ref{fig:model}. 
  \end{tablenotes}
\end{threeparttable}
\vspace{-.5em}
\end{table*}

\rev{We observed from the table that the only L1 concepts which exist in the reference model, yet were not identified in the literature are \textit{CONTROLLER REPRESENTATIVE} and \textit{PD PROVISION OBLIGED}, as shown in Table~\ref{tab:coverage}. The former refers to an administrative representative acting on behalf of the data controller,  while the latter concerns the denial of services when certain personal data are not shared with the controller.
From a research standpoint, these concepts seem not to be of 
particular relevance. 
A similar observation can be made with regards to missing L2 
and L3 privacy concepts (that are not listed in the table due to 
space limitations). 
}
\rev{These concepts are largely associated with requirements that are unlikely to be implementable; no software component is expected to address them and hence they are likely to receive little attention from the SE community. Nonetheless, such concepts typically entail legal obligations affecting the involved parties in the data collection and processing activities, including the mobile software development companies. } 
To be fully GDPR compliant, these concepts must be comprehensively specified in legal agreements, such as privacy policies or data processing agreements, established among the involved parties.

We observe that the literature is dominated by three \rev{privacy concepts}.
The most investigated concept (in 32 primary studies) is the \textit{PERSONAL DATA ORIGIN (L1)}, which refers to the \rev{source for the} personal data collection, e.g., \rev{either directly provided by the data subject (L2) or indirectly (L2), for instance,} through cookies (L3). 
The second is \textit{RECIPIENTS (L1)} (in 26 studies), which refers to the sharing of personal data with entities or organizations other than the controller, e.g., with third parties. 
Finally, \textit{LEGAL BASIS (L1)} is another well-investigated concept (in 21 studies), where the literature exclusively focus on \textit{CONSENT (L2)}: all of the 21 studies focus on \textit{CONSENT (L2)}; only one primary study had mentions of the remaining \textit{LEGAL BASES (L1)}.

We note that, while \textit{CONSENT} has a significant value since it authorizes a software system (or mobile app in this case) to process personal data of individuals, other legal bases such as \textit{LEGITIMATE INTEREST (L2)} can impact the implementation of certain actions related to users' data. For instance, the deletion of personal data upon the users' request can be postponed or declined because of the \textit{LEGAL OBLIGATION (L2)} based on which personal data was collected in the first place. Personal data can further be processed based on lawful \textit{PUBLIC INTEREST (L2)} (e.g., public health or historical research purposes). 

Another key observation is that, despite the significant attention given to \textit{CONSENT(L2)}---since it concerns individuals (i.e., users of a given mobile app)---very little attention has been given to \textit{(DATA SUBJECT RIGHT (L1)} (only in \rev{nine} primary studies, \rev{with three specifically on \textit{WITHDRAW CONSENT (L2)}}). 
\rev{\textit{DATA SUBJECT RIGHTS (L1)} concern} individuals' rights over their personal data, another user-centered concept that is highly-relevant for the development of legally-compliant mobile apps.
The primary studies remained at a generic level \rev{(L1) except for \textit{WITHDRAW CONSENT (L2)} (explicitly mentioned in three papers)} to enable the user to revoke their consent given in the first place.
Essential rights such as the right to \textit{ACCESS (L2)} where users should be able to access their personal data collected over a certain period of time, or the right to \textit{ERASURE (L2)} enabling users to request the deletion of their personal data were not specifically considered.
Allowing individuals to practice their rights is fundamental according to GDPR. 
If such legal requirements are well-captured and appropriately implemented in mobile apps, individuals can easily practice their rights through, e.g., direct interaction within the app.

\rev{Overall, we observe that, apart from \textit{CONSENT (L2)} and
  \textit{WITHDRAW CONSENT (L2)}, the majority of the studies discuss
  GDPR concepts at a high level (mostly L1), overlooking specific
  details (L2 and L3).} \rev{We note that a privacy
  concern being well-covered in the literature does not necessarily
  imply that it has been resolved. Rather, it reflects the growing
  interest in and awareness of the community of this concern. Conversely, under-explored concerns, or the lack of in-depth analysis, may indicate limited awareness, which could, in turn, lead to non-compliance practices. }

Considering that the most well-covered concept is \textit{PERSONAL DATA ORIGIN (L1)}, \rev{and \textit{PERSONAL DATA CATEGORY (L1)} also being well represented (13)}, we looked closely at the types of personal data \rev{(C2.1 in Table~\ref{tab:extraction})} that are being discussed in the primary studies.  
Broadly, there are three categories of personal data types \rev{(related to \textit{PERSONAL DATA CATEGORY (L1)}}. 
The first category concerns data about individuals, which are not app-dependent, i.e., they do not change depending on which app is being used. 
Examples in this category include name, phone number, home address, gender, age, bank details, health information, photos, and salary. 
The data in this case are often linked to the \textit{DIRECT (L2)} concept, i.e., personal information collected directly from the user of an app,  (mentioned in 30 primary studies out of the 32 discussing \textit{PERSONAL DATA ORIGIN (L1)}). 
The second category concerns data about the user of the mobile app, e.g., user name, password, phone calls, SMS, contacts. 
Such data often varies across different apps, depending on, for example, the domain and the purpose of the app.  
The last category concerns data related to the mobile device, e.g., geolocation information, IP address, connectivity status, SIM card ID, addresses of nearby hotspots. 
We believe it is important to make this distinction at the time of developing the app, to have a better understanding of what data will be collected and hence what requirements are necessary to fulfill in order to develop apps that comply with GDPR. 
We remind the reader that the GDPR imposes obligations for protecting personal data. 
It requires more strict obligations (e.g., security mechanisms) when sensitive personal data (e.g., health information) is collected compared to less or non sensitive data, such as  the user name invented  by an individual for using a specific mobile app.

\begin{tcolorbox}[colback=black!5!white,colframe=white!50!black,]
\rev{\textit{In response to RQ1,} we found that the literature mostly
  focuses on high-level concepts of the model and mostly three
  concepts: (1) PERSONAL DATA ORIGIN, especially the collection of
  personal data; (2) RECIPIENTS, which has to do with the sharing of
  personal data;  (3) LEGAL BASIS, and more specifically L2 CONSENT,
  which deals with the informed consent provided by the data subject. 
In contrast, several L2 concepts, such as WITHDRAW CONSENT or all the
DATA SUBJECT RIGHTS, which are essential concerns of the GDPR
regarding the protection of the data subjects, are not enough represented.}
\end{tcolorbox}

\subsection{RQ2. What are the main objectives pursued when
  investigating GDPR privacy concepts?}
\label{sec:rq2}

RQ2 investigates the main research problems related to GDPR compliance addressed by the SE community. The main objectives observed in our work can be summarized across eight themes (T1--T8), further discussed in the rest of the section: 
\begin{enumerate}[T1.]
	\item \rev{Tracing personal data in mobile apps}
    \item \rev{Analyzing app permissions}
    \item \rev{Analyzing the security and privacy of minors}
    \item \rev{Analyzing the implementation of informed consent}
    \item \rev{Analyzing users' awareness on security and privacy}
    \item \rev{Analyzing the consistency between mobile apps 
    and their privacy policies}
	\item \rev{Analyzing the implementation of the data minimization principle}
    \item Summarizing privacy policies
\end{enumerate}

Table~\ref{tab:rq2} lists the primary studies corresponding to each theme. 
While our analysis here focuses primarily on the objectives of the primary studies, we discuss below also the relevant \rev{concepts} under each theme to better connect with our findings in RQ1. 
We identified relevant \rev{concepts} using a deductive analysis. 
\rev{Recall from Section~\ref{subsec:dataextraction} that we adopted an \textit{integrated coding approach}. For this RQ, we first synthesized the themes using \textit{inductive analysis}, then we applied \textit{deductive analysis} to map themes to the pre-defined concepts derived from the \textit{reference model}. 
}

   \begin{table}
\centering
\caption{Overview of Themes Observed in the Literature (RQ2)}
\label{tab:rq2}

  \centering
   \begin{tabularx}{0.98\linewidth}{@{} p{0.12\linewidth} @{\hskip 0.5em} p{0.06\linewidth} *{1}{>{\arraybackslash}X}@{}}
     \toprule
    Theme  &  Total & Studies\\
    \midrule 
    T1     &   8    &   \citeS{S1,S28,S41,S46,S50,S51,S52,S59}   \\
    T2    &   6    &   \citeS{S2,S3,S20,S25,S27,S57}    \\
    T3     &   3    &   \citeS{S9,S24,S32}   \\
    T4    &   11   & \citeS{S5,S7,S11,S21,S39,S42,S44,S45,S48,S49,S56}    \\
    T5    &   10   &   \citeS{S16,S19,S22,S23,S26,S31,S33,S34,S36,S58}    \\
    T6    &   7    &   \citeS{S18,S29,S35,S37,S38,S55,S60}    \\
    T7     &   7    &   \citeS{S4,S8,S10,S12,S13,S14,S43}   \\
    T8    &   8    &   \citeS{S6,S15,S17,S30,S40,S47,S53,S54}    \\
    \bottomrule
 \end{tabularx}

\end{table}

\subsubsection{T1. \rev{Tracing personal data in mobile
    apps}}
\label{sec:t1.-revtr-pers}

Primary studies often trace certain personal data to identify potential data leakage, \rev{i.e., unintended or unauthorized data exposure beyond the mobile app, such as its sharing with third parties. This leakage may lead to unlawful disclosure and/or data breaches.}. 
This theme was observed in eight studies. 
Identifying which personal data to trace is a relevant research topic.

\rev{According to our \textit{reference model}, the categories of personal data being collected (\textit{PERSONAL DATA CATEGORY}}) must be clearly specified by the controller 
and explicitly communicated to the data subject (i.e., the users of these mobile apps) through a privacy policy that must be agreed prior to using the app. 
In other words, no personal data should be collected without having been declared in the privacy policies upfront. 

Fostering this research theme is essential for understanding whether the mobile app is GDPR compliant when collecting personal data from the user. The analysis of source or byte code of mobile apps should be better aligned with the GDPR provisions as well as what has been stated in their respective privacy policies.  

We believe that this theme crosses multiple \rev{concepts}, including \textit{PERSONAL DATA CATEGORY}, \textit{PD ORIGIN}, and \textit{CONSENT}. 
Sharing data with third-party libraries is a closely related theme; in our \textit{reference model}, this theme concerns the \rev{concept of} \textit{RECIPIENT}. 
In brief, the GDPR requires all recipients to be listed in the privacy policies of mobile apps. If the app is sharing personal data with third-party libraries, then this has to be fully transparent. 

\subsubsection{T2. \rev{Analyzing app permissions}} 

This theme investigates the access to personal data granted to the mobile app through permission requests. 
As mobile apps are becoming more pervasive, this theme is important
for understanding what personal information the mobile app can access
or collect indirectly through permission requests, e.g., to access
information in other apps, \rev{risking that apps may gain access to
  data they were originally not allowed to access}. 
Typical example is when the user ought to share location information to be able to use various apps such as navigation or fitness apps.
This theme is related to the following \rev{concepts}: \textit{PERSONAL DATA SECURITY}, \textit{RECIPIENTS} and \textit{PD ORIGIN}, the latter two being among the most frequent \rev{concepts} as per RQ1. 
We observed this theme in six primary studies.

\subsubsection{T3. \rev{Analyzing the security and privacy of minors}} 
\label{sec:t3.-revan-secur}
This theme focuses on minors, with an emphasis on the threats, risks, and \rev{challenges pertinent to this user group (e.g., obtaining valid informed consent), and the need to  ensure  their rights are adequately protected. }
This theme maps directly to the \textit{CHILDREN} \rev{concept}. We observed this theme in three primary studies.  

\subsubsection{T4. \rev{Analyzing the implementation of informed consent}} 

Consent is a well-studied concept. The GDPR emphasizes on collecting explicit consent from individuals, enabling them to make informed decisions with regard to the handling of their personal data. However, according to the \textit{reference model} we apply in this SLR, \textit{CONSENT} is only one of the several possible legal bases for data collection and processing (occurring in 32 studies). 

\textit{CONSENT} is also the easiest to verify in software as it should be translated into interaction between the mobile app and the user. To investigate whether the practices in the mobile app code are GDPR compliant with respect to some other legal bases,  a more comprehensive analysis of the app context must be conducted; such an analysis would not be necessarily limited to, e.g., the app's source code. We note that, despite not being popular in the SE literature, legal bases other than consent (which are mentioned in only one paper out of 32) can play an important role in mobile apps development decisions. 

\subsubsection{T5. \rev{Analyzing users' awareness on security and privacy}}

This theme (observed in ten studies) surveys the users of mobile apps to study how well individuals are aware of their rights as well as  common practices concerning privacy and security issues, often perceived in mobile apps' behaviors. While this theme highlights the importance of individuals' awareness about data protection, it cannot be directly mapped to our \textit{reference model}.

\subsubsection{T6. \rev{Analyzing the consistency between mobile apps and their privacy policies}} 

Some papers investigate the consistency between the content of the privacy policy associated with an app against the actual behavior of that app. However, this research theme is often scoped to a use case such as the collection and sharing of sensitive personal data with third parties. Seven studies deal with this theme.

As pinpointed earlier, this theme is important to ensure that mobile apps are compliant with what is disclosed in their privacy policies, upon which individuals agree. 

Advancing this theme requires in-depth analysis of at least the following \rev{concepts}: \textit{PERSONAL DATA CATEGORY}, \textit{LEGAL BASIS}, \textit{RECIPIENTS}, \textit{PD ORIGIN}, and \textit{DATA SUBJECT RIGHT}. These \rev{concepts are related to} the understanding of what personal data is collected, the source where it is obtained from, with whom it will be shared, what rights individuals can have on their personal data, and how these rights are implemented.  

\subsubsection{T7. \rev{Analyzing the implementation of the data minimization principle}} 

Data minimization is an important guiding principle in GDPR: according to this principle, an app should not collect more personal data than necessary for its use. We observed this theme in seven primary studies.

Such a theme would require analyzing the context of the mobile app, specifically the \textit{LEGAL BASIS} under which personal data is being collected and processed as well as the \textit{PROCESSING PURPOSES} declared in the privacy policies of mobile apps. 

To ensure that no personal data is unnecessarily collected, the findings related to these two \rev{concepts} should then be cross-checked against \textit{PERSONAL DATA CATEGORY}, indicating the personal data types collected, and \textit{PD ORIGIN}, indicating the source where the data is collected from. 

\subsubsection{T8. Summarizing privacy policies} \label{sec:t8.-summ-priv}

This theme (observed in eight studies) focuses on making privacy policies more accessible to \rev{individuals (e.g., end-users or engineers) } by providing informative summaries or categorization of the policies. 

While the primary studies under this theme cannot be one-to-one mapped to \rev{concepts} in our \textit{reference model}, we believe that the \textit{reference model} can  be useful to advance this research theme as it provides a structured hierarchy of the necessary details that must be identified in privacy policies.

\noindent 
\rev{In brief, this study aims to provide a comprehensive overview of the literature landscape addressing GDPR privacy concerns in mobile app research. While RQ1 serves as a starting point by mapping existing work to the concepts defined in our reference model, RQ2 examines the main objectives pursued by the primary studies. The rationale for linking RQ2 back to RQ1 is to offer a deeper insight into why certain concepts were addressed and others were not, driven by the overarching objectives of the primary studies. }

\begin{tcolorbox}[colback=black!5!white,colframe=white!50!black,]
\rev{\textit{In response to RQ2,} we identified eight main themes in the literature, including tracing personal data for detecting potential data leakage, analyzing consent implementation and data minimization principles, checking the consistency between granted permissions, data use, and privacy policies, as well as topics addressing the privacy of minors, users' privacy awareness, and the summarization of privacy policies.}
\end{tcolorbox}


\subsection{RQ3. What are the different types of research contributions?}
\label{sec:rq3}
RQ3 explores the various contributions identified in the literature concerning GDPR.
We distinguish in this SLR between \textit{technical} and 
\textit{non-technical} contributions: the former refers to  
\rev{automated or semi-automated approaches proposed to  
address specific GDPR-related privacy concerns}, 
\rev{whereas the latter focus on sharing insights or 
recommendations, often derived from surveys, with the goal of 
advancing the research knowledge in this domain.}  
\revv{Additional contextual information extracted from the 
primary
studies, including application domains and targeted mobile
platforms where available, is provided in the supplementary 
material accompanying this paper~\cite{annex}.}

In total, 35 out of 60 primary studies have non-technical contributions, targeting various GDPR privacy concerns. 
The majority of them conducted empirical studies as well as literature or apps surveys to learn more about mobile apps' users and practices in different contexts. 
Concrete outcomes of such studies include, for instance, lessons learned and guidelines on the development of GDPR-compliant mobile apps (which also address users' privacy concerns). 

\rev{Out of these studies, 34} are qualitative, relying on manual means for analyzing certain privacy concerns. However, some studies used simple tool support such as keyword-based or string search for extracting indicator terms, or applied third-party tools such as Mobile Security Framework (MobSF)\footnote{\url{https://github.com/MobSF/Mobile-Security-Framework-MobSF}}, FlowDroid\footnote{\url{https://github.com/secure-software-engineering/FlowDroid}} for static or dynamic code analysis, \rev{or Wireshark\footnote{\url{https://www.wireshark.org/}} for network traffic analysis}. 
Such an automation was strictly used to support their 
empirical studies or surveys. 

\rev{As mentioned previously,} most primary studies with non-technical contribution often described user studies or app surveys. 
The number of participants involved in user studies performed under this category vary between eight to 6,124  participants.   
Similarly, the number of apps considered in apps surveys vary, 
ranging from small app datasets (27 in~\citeS{S3}, or 28 
in~\citeS{S2}) to more than 10,000 apps in large-scale studies 
such as~\citeS{S12}. 

The remaining 25 primary studies make technical contributions. 
They proposed \rev{automated or semi-automated approaches, regardless
  of whether they have resulted in full-fledged automated tools or prototypes. } These approaches leverage different enabling technologies: model-driven engineering was used to enhancing business process modeling notation with privacy considerations~\citeS{S6},   
static and dynamic program analysis were used for detecting 
violations or vulnerabilities in mobile 
code~\cite{S35,S41,S46,S51,S59},  
NLP was employed for analyzing privacy policies and app 
reviews~\cite{S36,S44,S53,S57}, machine learning (ML) and 
deep learning (DL)
were used for classifying privacy policy provisions~\cite{S30} 
and \rev{taint analysis for detecting potential data leaks in app code}~\cite{S1}.    
\rev{13 of the primary studies with technical contributions } leveraged 
a combination of these enabling  technologies, particularly NLP, ML, DL, and static or dynamic program analysis. 
These studies pursued the following objectives: 
\begin{itemize}
\item detecting privacy issues in app reviews~\cite{S16}, 
\item identifying data recipients~\cite{S8}, 
\item detecting sensitive information disclosure in mobile apps~\cite{S20},
\item addressing privacy considerations entailed from using third-party apps or libraries~\cite{S28,S47},
\item evaluating the consistency between privacy policies and actual app code~\cite{S37,S52,S54,S60}, 
\item detecting abuse of data collection due to granted permissions~\cite{S42,S43,S49,S50}
\end{itemize}

The aforementioned technical contributions were in all cases accompanied with validation elements such as, e.g., proof of concept examples~\cite{S6}, user studies~\cite{S16,S59}, or empirical validation using datasets.
These datasets vary in size from small (with less than 50 elements such as apps, privacy policies, app reviews, or programming methods~\cite{S42}),  medium (51--500 elements~\cite{S16,S28,S30,S50,S52,S53,S57}), to large (with more than 500 elements~\cite{S1,S8,S20,S35,S36,S37,S41,S44,S46,S49,S51,S53,S54}.)

As part of this RQ, we also looked into the app stores and 
operating systems targeted in the literature. Our results show 
that \rev{32} papers 
considered Android only, \rev{seven} 
considered both Android and iOS, one paper considered Android and Windows systems, and one more paper focused exclusively on iOS.
\rev{The literature aligns with Android being the largest share of the mobile market compared to iOS. 
Studies covering both operating systems often rely on user surveys~\cite{S11,S19,S9}, governance or policy analyses~\cite{S10,S18,S38}, third-party consent management platforms~\cite{S49}, or screenshots of user interfaces~\cite{S57}, rather than focusing on concrete app code analyses tasks in relation to GDPR-related privacy concerns.}

By examining the 25 studies with technical contributions, we observe that all but four are exclusively focused on Android.
Regarding the outliers, one utilizes Amazon Alexa (a virtual assistant)~\cite{S37}, one is based on iOS~\cite{S59}, and the last two mention Apple App Store as well as Google Play Store for work based on explicit consent (through GUI testing) and privacy policies~\cite{S49,S57}.
The remaining 21 studies focusing on Android include: One study that investigates  Xiaomi market Store~\cite{S43}, five lack comprehensive details about app stores~\cite{S41,S42,S44,S50,S60}, and the remaining ones utilize GooglePlay~\cite{S8,S16,S20,S28,S30,S35,S36,S41,S44,S46,S47,S51,S52,S53,S54}.

\begin{tcolorbox}[colback=black!5!white,colframe=white!50!black,breakable]
	\rev{\textit{In response to RQ3,} we found that research
          contributions can be broadly categorized into two types:
          technical contributions, focusing on providing (semi-) automated support to address specific privacy concerns, and non-technical contributions, which are mostly based on user surveys. Most existing research focuses on the Android market and, to a limited extent, iOS. 
		We also observed that several privacy concerns remain under-explored in the literature, calling for further research. }

\end{tcolorbox}


\subsection{RQ4. Are the research artifacts accompanying studies on GDPR privacy concepts publicly available?}

Out of the 60 papers, only 14 papers ($\approx$23\%)  have shared some artifacts associated with the  conducted research;
Table~\ref{tab:material} lists these papers and indicates the type of artifacts (source code, datasets, evaluation data files) made available. Out of these 14 papers, only two papers~\cite{S37,S50} released all three types of artifacts. 

Table~\ref{tab:license} further investigates the licensing schema and on which type of platform is located the available material.
We note that two studies~\cite{S37,S46} have different licensing schemes regarding their source code and the additional material, which are stored separately in a different repository.

Regarding licenses, 5 of the 14 papers do not mention any explicit license and an additional 6th consider a vague license (other - open) leading to $\approx$43\% of the work not being explicit about their licensing schemes.
Regarding repositories, 11 of the 14 papers share (part of) their material through project webpages on personal website or GitHub repositories; only five consider persistent third-party repositories, including one study on the publisher's page for additional content and four studies on Zenodo (with two papers sharing material in both Github and Zenodo).

These results show that the open science practices in the SE community still need improvement in terms of adding appropriate licenses and sharing material through persistent repositories such as Zenodo or FigShare.

\begin{table}
\centering
\caption{Publicly Available Material}
\label{tab:material}
  \centering

 \begin{tabularx}{0.95\linewidth}{@{} p{0.14\linewidth} @{\hskip 0.5em} p{0.14\linewidth} @{\hskip 0.5em} p{0.14\linewidth} *{1}{>{\arraybackslash}X}@{}}
     \toprule
    Source Code & Datasets & Results & Papers \\
    \midrule
       \ding{51}  &  \ding{51}&  \ding{51}   &   \citeS{S37,S50} \\
       \ding{51}  &  \ding{51}&       \ding{55}        &   \citeS{S28,S46,S47} \\
       \ding{55}       &  \ding{51}&  \ding{51}   &   \citeS{S2} \\
       \ding{51}  &       \ding{55}     &       \ding{55}        &   \citeS{S1,S43,S48,S53,S60} \\
       \ding{55}       &  \ding{51}&       \ding{55}        &   \citeS{S8,S21} \\
       \ding{55}       &       \ding{55}     &  \ding{51}   &   \citeS{S3} \\
    
    \bottomrule

 \end{tabularx}
\end{table}

\begin{table}
\centering
\caption{Licensing and Sharing Approaches over Available Material}
\label{tab:license}
  \centering
 \begin{tabularx}{0.95\linewidth}{@{} p{0.4\linewidth} @{\hskip 0.5em} p{0.1\linewidth} *{1}{>{\arraybackslash}X}@{}}
     \toprule
    
    License type & Total & Papers \\
    \midrule
    No license mentioned & 5 & \cite{S2,S21,S28,S48,S50}\\
    Apache 2.0 & 1 & \cite{S37}\\
    CC0-1.0 license & 1 & \cite{S46}\\
    Creative Commons Attribution 4.0 International & 4 & \cite{S3, S46,S47,S53}\\
    CC BY 4.0 & 1 &  \cite{S8}\\
    GNU GPL 3.0 & 2 & \cite{S1,S60}  \\
    MIT License & 1 & \cite{S43} \\
    other (open) & 1 & \cite{S37}\\
    \toprule
    Sharing type & Total & Papers \\
    \midrule
    Personal website & 1 & \cite{S28}\\
    GitHub Project & 10 & \cite{S1,S2, S21,S37,S43,S46,S48,S50,S53,S60}\\
    Publisher Repository & 1 & \cite{S3} \\
    Open Science Repository & 4 & \cite{S8,S37,S46,S47} \\
    
    \bottomrule

 \end{tabularx}
\end{table}

\subsection{Research Gaps}
\label{subsec:gaps}

Our findings indicate that despite the attention received by GDPR in mobile app research, existing work focuses only on certain aspects, such as consent or sharing data with third-party libraries. Leveraging the reference model by Amaral et al.~\cite{Amaral:21}, we identify  research gaps that need yet to be covered, outlined below.  

\textbf{G1. More research is needed to better understand  
personal data collection practices in mobile apps. } 
Table~\ref{tab:coverage} shows that the primary studies have extensively investigated the collection of personal data and recipients of personal data sharing activities. These topics are often related to analyzing security issues concerning data leaks. To this end, we identify the following  research gaps:

\begin{itemize}
    \item[G1.1] While there has been some work toward 
    discovering sensitive personal data collected and processed 
    in mobile apps, \revv{and recent work has examined 
    compliance with
    privacy regulations such as the CCPA~\cite{samarin2023}, 
    there remains
    considerable room for further research on tracing the 
    personal
    data that is actually collected, processed, and shared by
    mobile apps in practice.  }
    According to Amaral et. al~\cite{Amaral:21}, privacy policies 
    of mobile apps should explicitly list the personal data 
    categories that are planned to be collected by the mobile 
    app;  sometimes the policy should also explicitly mention the 
    origin from where such personal data is being collected. 
    \textit{Verifying  whether the categories mentioned in the 
    privacy policies are also the only ones being collected and 
    processed by mobile apps is essential for checking the 
    compliance of apps. }
    
    \item[G1.2] Nowadays, personal data collection typically 
    involves permission requests to access personal data 
    collected by other apps on the same mobile device. 
    \textit{Another research gap concerns the investigation of 
    indirect collection of  personal data.} It is important to 
    understand what personal data is collected indirectly (e.g., 
    scraping the web for data  available online sources or 
    accessing datasets through data brokers), how the process 
    is documented, and whether users are aware of that.
    
    \item[G1.3] \rev{The literature has carefully investigated 
    personal data categories that are directly collected from 
    individuals based on, e.g., consent. Personal data that 
    describe users or their mobile devices,  which are typically 
    collected indirectly through other means are not very well 
    explored in the literature.
	\textit{A key research gap is to investigate to what extent such  personal data categories contain personal identifiable information and would therefore require special attention concerning GDPR compliance practices.} For instance, the mobile IP address could reveal the geographical location of an individual. }. 
\end{itemize}

\textbf{G2. Further research is needed to investigate legal bases beyond ``consent'' and their impact on the development of legally-compliant mobile apps.}
Our results show that most primary studies investigate the explicit consent of a user for data collection and processing. While consent is indeed an important legal basis, other legal bases are as important. For instance, collecting and processing personal data based on legitimate interest for the purpose of  investigating a security attack is a good example where consent might not even be asked. \textit{The legal basis for handling personal data provides a complex, yet unexplored research landscape.} 

\textbf{G3.  There is a need for further research on the 
elicitation of comprehensive, implementable requirements 
pertaining to data subject rights. }
Our SLR highlights that the existing work investigated data 
subject rights and their implementation in mobile apps only to 
a very limited extent. 
According to the GDPR, users (as data subjects) should be 
able to easily exercise their rights to access their personal 
data, lodge a complaint, etc. For instance, mobile apps should 
provide adequate and explicit information (ideally through 
direct interactions) for users to request account deletion or to 
access all their personal data collected during a specific time 
span~\cite{EDPB2022}.  \textit{In-depth understanding of data 
subject rights should  first be established in research to then 
make its way into practice}. 
Without such an awareness level on the different users' rights,
developing better and legally-compliant apps is  subject to the
random, potentially correct interpretations of developers. 

\section{Threats to Validity}
\label{sec:threats}

\sectopic{Internal Validity.} The main consideration concerning the internal validity is selection bias. To mitigate the effect of this threat, we followed the guidelines reported in the literature for conducting systematic literature reviews. The results of our data extraction are also made publicly available and are thus open to scrutiny. 
\revv{The fact that the snowballing process did not
  identify additional eligible studies provided further confidence
  in the coverage of the search strategy. Nevertheless, some
  relevant studies may have been missed due to terminology
  variations, indexing limitations, or database coverage. The
  results of our data extraction are publicly available and thus
  open to scrutiny and replication.}
\rev{Another potential threat stems from the absence of a 
formal quality assessment
	checklist, which might have resulted in unequal treatment of 
	primary studies. To mitigate this
	threat, we applied strict inclusion criteria based on the topic
	relevance, and we held several meetings among the authors 
	to ensure shared understanding and minimize subjective 
	judgment.}

\sectopic{Construct Validity.} The primary threat to construct validity is related to data extraction. To minimize this threat, the first three authors have conducted several online sessions to discuss the findings and ensure the mutual understanding of the extracted information categories. We further computed the inter-rater agreement on the mapping activity that was primarily performed by the first author, to ensure that the GDPR privacy concerns analyzed in this SLR are correctly mapped to the main topics discussed in the primary studies. 

\sectopic{External Validity.}
\rev{The findings of this study are primarily pertinent to GDPR-relevant privacy concerns in mobile app research, which may limit their generalizability to other regulations or software systems. We note that several identified concepts, such as informed consent, and corresponding research gaps, e.g., the need to explore legal bases beyond consent, may also be relevant across other research areas. However, we acknowledge that we provide no empirical evidence on their applicability beyond the scope of our study. 
To partially mitigate this threat, we explicitly defined the time range of interest for our SLR on the GDPR, iteratively refined our search query,  and retrieved the primary studies from multiple online repositories and search engines. We further interpreted the results with caution and grounded our findings in existing research. We leave it to the research community to examine whether similar patterns can be identified in broader research contexts.}

\revv{\sectopic{Conclusion Validity.} 
 The conclusions of this study may be influenced by the choice 
 of
the conceptual framework used to classify the primary studies.
To mitigate this threat, we adopted the model proposed by
Amaral et al.~\cite{Amaral:21}, which was developed in collaboration with legal
experts and is grounded in GDPR provisions. Although the 
model
was originally applied to privacy-policy analysis, the concepts
it defines represent broader GDPR-related privacy concerns 
that
extend beyond privacy policies and are applicable to the study
of mobile applications.}

\section{Conclusion}
\label{sec:conclusion}

In this paper we have presented our systematic literature 
review of the research landscape in software engineering (SE) 
on addressing GDPR-relevant privacy concerns in mobile 
apps. We have  reviewed 60 primary studies and analyzed 
them with respect to an existing conceptual model that 
\rev{provides a comprehensive set of concepts representing 
the GDPR.} 

\rev{We found that existing studies predominantly address three key GDPR-related privacy concepts: (i) the collection of personal data directly from users through mechanisms like cookies \textit{PD ORIGIN}, (ii) the sharing of personal data with external entities (e.g., third parties) beyond the mobile apps (\textit{RECIPIENTS}), and (iii) the acquisition of consent from users (\textit{LEGAL BASIS - CONSENT})}. 
Additionally, our results show that the majority of existing work investigates problems such as  data leakage, app permission requests, and obtaining users' consent. To address these challenges, existing work relies to a large extent on natural language processing, machine learning, and program analysis.

\rev{We further identified research gaps 
that highlight the need for further research in: (i) personal data collection, involving both checking the consistency between personal data categories disclosed in the app's privacy policy against what the app actually collects as well as personal data collected indirectly, (ii) investigating the impact of legal bases beyond consent on the development of compliant mobile apps, and (iii)
eliciting detailed implementable requirements pertinent to data subject rights. }

\section*{Acknowledgment}

This research was funded in whole, or in part, by the Luxembourg National Research Fund (FNR), grant reference NCER22/IS/16570468/NCER-FT.

\bibliographystyle{elsarticle-num} 
\bibliography{paper}

\bibliographystyleS{elsarticle-num} 
\bibliographyS{studies.bib}

\end{document}